\documentclass[fleqn,12pt,twoside]{article}
\usepackage{espcrc1}

\usepackage{graphicx}

\title{Coherent Pion Photoproduction on Deuterium }

\author{Y. Ilieva\address[GWU]{The George Washington University/Jefferson Lab, \\
        12000 Jefferson Avenue, \\ 
        Newport News, VA 23606, USA}%
        \thanks{This work was supported by the U.S. Department of
                Energy under grant DE-FG02-95ER40901. Southeastern Universities
                Research Association (SURA) operates the Thomas Jefferson National
                Accelerator Facility under U.S. Department of Energy contract DE-AC05-84ER40150.},
        for the CLAS Collaboration}

\begin{document}

\maketitle

\begin{abstract}
Differential cross sections of the $\gamma d\rightarrow d\pi^0$ reaction were measured at
photon beam energies between 0.5 and 2.0 GeV over a wide angular range in the CM system. The
experiment was performed in Hall B at Jefferson Lab using the CLAS detector. The excitation
function, especially at backward CM scattering angles, has been studied with regard to
effects due to intermediate on-shell $\eta$ and $\omega$ rescattering. At CM scattering
angles greater than 123$^\circ$ our data clearly show an enhancement around 0.7 GeV,
which could be attributed to an intermediate on-shell $\eta$-state.
\end{abstract}

\section{INTRODUCTION}
In a previous measurement \cite{Ima85} of the reaction $\gamma d\rightarrow d\pi^0$, 
covering several backward-pion CM scattering angles and photon
energies between 0.5 and 1.0 GeV, a structure has been observed in the
excitation function at $\theta^{cm}_{\pi}$=130$^\circ$ between 0.7 and 0.8 GeV which
could not be explained in terms of non-exotic mechanisms \cite{Miy}. However, an approach
including an intermediate on-shell $\eta$-meson rescattering (which has been successful
in describing a similar feature of the $\pi d$ elastic backward excitation function 
\cite{Kon}) might be able to account for such an enhancement around the $\eta$-photoproduction
threshold. An intermediate on-shell $\omega$ formation should appear in the
$\gamma d\rightarrow d\pi^0$ excitation function close to the $\omega$-photoproduction
threshold \cite{Stra}. Thus, if the finding in \cite{Ima85} is confirmed,
a detailed study of this process extended to higher beam energies and broader
angular range may prove to be a valuable tool for studying $\eta$N and $\omega$N
interactions, which is the goal of our efforts.


\section{EXPERIMENT}
The experiment was performed in Hall B at Jefferson Lab using the CLAS detection
system \cite{CLAS03}. Electrons from the CEBAF accelerator hit a thin gold radiator
and produced a bremsstrahlung beam that was then incident on a 10-cm thick liquid-deuterium
target. The energies of the incoming
photons were measured directly with the Hall-B Photon Tagger \cite{Sob00}, whereas the
reaction products were detected using CLAS. The intensity of the 
photon beam was $\sim$10$^{7}$ tagged photons/s and the generated luminosity was around
$4\times10^{30}$ s$^{-1}$cm$^{-2}$. Data were obtained for electron energies of 2.5 and 3.1 GeV, which
corresponds to photon energy ranges of 0.5-2.4 and 0.8-2.9 GeV. The number of
collected $d\pi^0$ events was $\sim$$1.8\times10^5$ and $\sim$$2.8\times10^4$ for the
two electron beam energies, respectively.

Since CLAS is very well suited for detecting charged particles and has
limited acceptance for neutrals,  
our analysis is based on detecting the deuteron and applying the missing-mass
technique to identify the reaction. 
The data were divided into 50-MeV-wide
photon-energy bins and 0.1- or 0.2- (depending on the statistics) wide 
cos$\theta^{cm}_{d}$ bins. For each energy and angular bin, background subtraction
was done by fitting the background in the deuteron missing-mass-squared distribution and 
subtracting it from the total yield. In order to determine the detector
acceptance, 10$^6$ phase-space-distributed $d\pi^0$ events were
generated and processed through the GEANT-based simulation code GSIM \cite{gsim} 
and the same analysis chain as the data. The acceptance varies between 40$\%$ and 70$\%$,
depending on the beam energy and CM scattering angle. Roughly estimated, the acceptance 
systematic uncertainty is at the level of 10$\%$.
Preliminary cross sections
were extracted by normalizing the yields to the photon flux and the number of target 
nuclei, and applying an acceptance correction. 

\section{RESULTS}
After analyzing the 2.5-GeV data set, we were able to extract preliminary
differential cross sections up to a photon energy of 1.2 GeV for backward-pion
CM scattering angles and up to 2.0 GeV for forward angles. The most backward-pion
CM scattering angle accessible in our data is 148$^\circ$ and is defined by the forward hole
in CLAS. The most forward accessible CM angle varies between 32$^\circ$ and 76$^\circ$,
depending on the beam energy, and is limited
by the detector energy acceptance for deuterons.

\begin{figure}[htb]
\begin{center}
\includegraphics[scale=0.6]{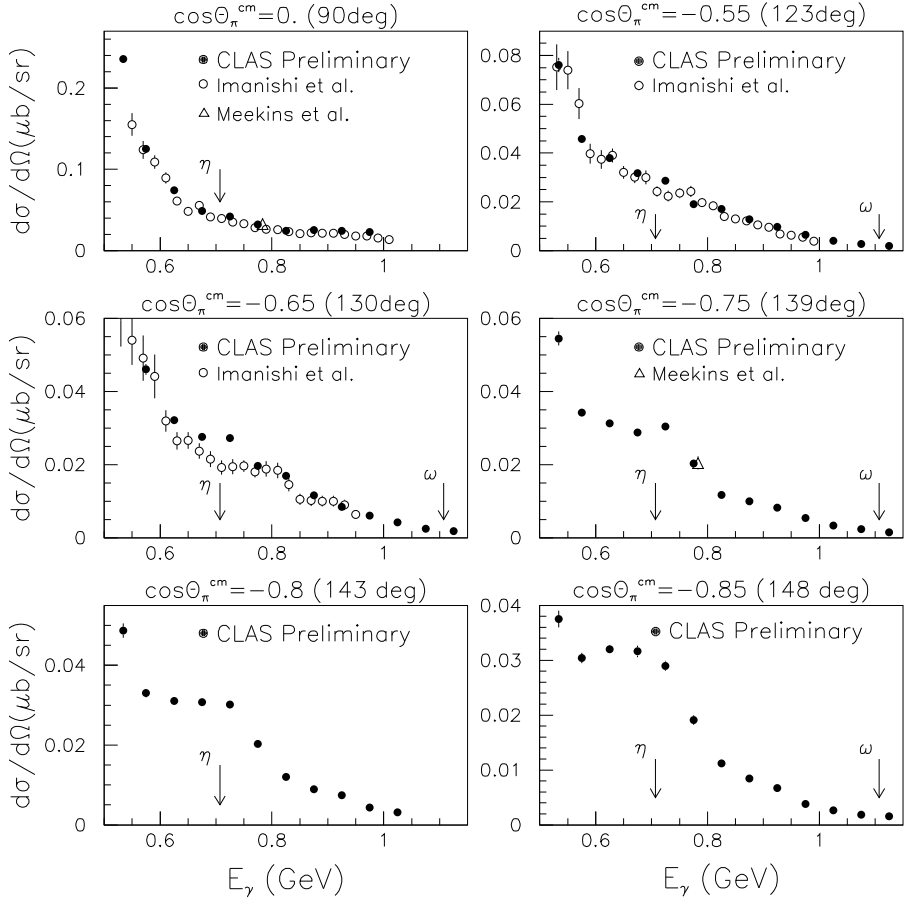}
\end{center}
\caption{Preliminary differential cross sections for the reaction $\gamma d\rightarrow d\pi^0$ 
obtained with the CLAS detector (solid symbols). Previous data \cite{Ima85,Meek99} are also shown 
(open symbols). The $\eta$- and $\omega$-photoproduction thresholds are indicated by arrows.}
\label{fig:yilieva_fig1}
\end{figure}

Our preliminary results for several pion CM scattering angles, compared with previous
measurements, are presented in Fig. \ref{fig:yilieva_fig1}. 
The uncertainties shown are statistical (3-12$\%$) and systematic due to 
background subtraction (up to 3$\%$) added in quadrature.
Fairly good agreement with the previous data is observed. All excitation
functions corresponding to $\theta^{cm}_{\pi}>$123$^\circ$ show a structure 
around the $\eta$-production threshold,
which becomes more prominent as the CM angle increases. This confirms the finding
reported in \cite{Ima85}. However, the shape of the structure we see is somewhat
different compared with the results of the previous measurement: at $\theta^{cm}_{\pi}$=130$^\circ$ our data
show an enhancement around 0.7 GeV, whereas the data from \cite{Ima85} show a broad plateau between
0.7 and 0.8 GeV. In order to understand this difference we are currently in the process 
of a detailed analysis of the systematics of our 
experiment. Due to statistics limitations, we might not be able to study the effect
of an intermediate on-shell $\omega$-state. 

\section{SUMMARY}
The reaction $\gamma d\rightarrow d\pi^0$ was measured at photon beam energies above 0.5 GeV. 
This is the
first detailed measurement of that process spanning wide ranges in energy and angle.
At photon energies and CM angles where there are existing data, our preliminary cross
sections are in good agreement with the previous measurements \cite{Ima85,Meek99}. We confirm 
the existence of a structure in the excitation function at $\theta^{cm}_{\pi}$=130$^\circ$
close to the $\eta$-photoproduction threshold. Moreover, our data extend to larger
CM angles where the structure is clearly more prominent. An approach including intermediate on-shell
$\eta$-meson formation \cite{Kon} will be used to describe the data in order to study the properties of the
$\eta$N interaction.


\begin{thebibliography}{9}
\bibitem{Ima85} A. Imanishi \textit{et al.}, Phys. Rev. Lett. 54 (1985) 2497
\bibitem{Miy} T. Miyachi and H. Tezuka, Phys. Rev. C 36 (1987) 844
\bibitem{Kon} L.A. Kondratyuk and F.M. Lev, Sov. J. Nucl. Phys. 23 (1976) 556 
\bibitem{Stra} I. Strakovsky, private communication
\bibitem{CLAS03} B. Mecking \textit{et al.}, Nucl. Instr. and Meth. A 503 (2003) 513
\bibitem{Sob00} D.I. Sober \textit{et al.},  Nucl. Instr. and Meth. A 440 (2000) 263 
\bibitem{gsim} E. Wollin, GSIM User's Guide, Version 1.1 (1996) 
\bibitem{Meek99} D.G. Meekins \textit{et al.}, Phys. Rev. C 60 (1999) 052201
\end{thebibliography}
\end{document}